\documentclass[aps,twocolumn,pra]{revtex4}
\usepackage{graphicx}

\begin{document}

\title{Extraction of information from single quanta}


\author{G.~S.~Paraoanu}
\affiliation{Low Temperature Laboratory, Aalto University, P. O. Box 15100, FI-00076 AALTO, Finland.}
%


\begin{abstract}
 We investigate the possibility of performing quantum tomography on a single qubit with generalized partial measurements and the technique of measurement reversal. Using concepts from statistical decision theory we prove that, somewhat surprisingly, no information can be obtained using this scheme. It is shown that, irrespective of the measurement technique used, extraction of information from single quanta is at odds with other general principles of quantum physics.
\end{abstract}

\maketitle

In a paper published 75 years ago \cite{epr}, Einstein, Podolsky and Rosen (EPR) formulated their famous criterion for elements of reality as follows: "if, without in any way disturbing a system, we can predict with certainty [...] the value of a physical quantity, then there exists an element of physical reality corresponding to this physical quantity". Is the wavefunction of a single quantum system an element of reality? A positive answer would entitle the wavefunction to an "ontological" status, much like Schr\"odinger believed to be the case, and in contradiction with the "epistemological" role reserved by the standard Copenhagen interpretation. Clearly, to think about the wavefunction as real, we would have to be able to measure it on a single quantum system. The question of whether this is possible was first raised in the context of  the so-called "protective" (weakly-disturbing) measurements in the early 1990', where it was answered in the negative \cite{protective}. The Copenhagen-school view of the wavefunction as a mere mathematical tool for calculating probabilities was saved.


However, protective measurements are not the only possibility. A different idea for measuring the wavefunction is to employ reversible positive operator valued measure
(POVM) measurements \cite{royer}. With the recent demonstration of reversibility of the  so-called "partial measurements"  in systems of phase qubits \cite{partialmartinis,katz} this idea  looks theoretically attractive and experimentally feasible. Here we will explore this strategy and consider
generalized partial measurements \cite{paraoanu}, which have the property that they can be probabilistically reversed for both results of the measurement. We then consider a series of measurements followed by reversals and we address the question of whether in this way it is possible to extract (with a certain success probability) any information about the qubit. We show by employing concepts from statistical decision theory that this cannot be done - all the information we get from the measurements is nullified by the very process of undoing them. Therefore, we cannot measure the wavefunction of a single quanta, and as such we are not entitled to regard it as an element of  reality. We further connect this result to more general physical principles, by examining the consequences of being able to perform quantum tomography on a single qubit for experiments such as EPR, quantum teleportation, quantum cloning, and quantum key distribution.

For consistency, we first briefly review the properties of generalized partial measurements \cite{paraoanu} for a qubit in a basis with states $|0\rangle$ and $|1\rangle$. We define two measurement operators, $M_{m}$ and $M_{\bar{m}}$, corresponding to measurement results "$m$" and "$\bar{m}$", and parametrized by two real numbers $p$ and $q$, $0\leq p,q\leq 1$,
\begin{eqnarray}
M_{m} &=& \sqrt{1-q}|0\rangle\langle 0| + \sqrt{1-p}|1\rangle\langle 1| , \label{embarg} \\
M_{\bar{m}} &=& \sqrt{q}|0\rangle\langle 0| + \sqrt{p}|1\rangle\langle 1| , \label{mg}
\end{eqnarray}
which implement a POVM measurement with effects (elements) $E_{m}=M_{m}^{\dagger}M_{m}$ and $E_{\bar{m}}=M_{\bar{m}}^{\dagger}M_{\bar{m}}$ \cite{nielsen}. If the qubit is in an unknown pure state $|\psi\rangle$, the probability of obtaining the result "$m$" is $P_{m} =\langle \psi |E_{m}|\psi\rangle$, and the probability of obtaining the result "$\bar{m}$" is
$P_{\bar{m}}= \langle \psi |E_{\bar{m}}|\psi\rangle$. The state after the measurement is $|\psi_{m}\rangle = (1/\sqrt{P_{m}})M_{m}|\psi\rangle $ in  the first case, and $|\psi_{\bar{m}}\rangle = (1/\sqrt{P_{\bar{m}}})M_{\bar{m}}|\psi\rangle$ in the second. The physical meaning of the parameters $p$ and $q$ is that of probabilities for a qubit in the state $|1\rangle$ respectively $|0\rangle$ to yield the result $\bar{m}$; in the case of Josephson-junction qubits these can be directly related to switching-current probabilities \cite{paraoanu,me}.

\begin{figure}[t]
\begin{center}
  \includegraphics[width=7.5cm]{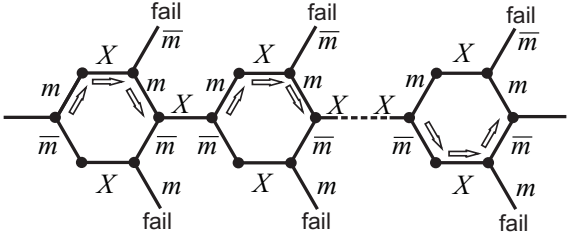}
\end{center}
\caption{Schematic of a series of generalized partial measurements and their reversal. The arrows indicate possible measurement-reversal paths actually occuring.}
\label{efig}
\end{figure}

Generalized partial measurements can be probabilistically reversed no matter which result, "$m$" or "$\bar{m}$", occurs under a measurement. The reversal is non-deterministic (probabilistic), in the sense that in both cases the reversal operation can also fail. More precisely, if $p$ and $q$ are neither 0 or 1, the operators $M_{m}$ and $M_{\bar{m}}$ can be inverted,
\begin{eqnarray}
M_{m}^{-1}&=& \frac{1}{\sqrt{(1-p)(1-q)}}XM_{m}X , \label{inv1}\\
M_{\bar{m}}^{-1} &=& \frac{1}{\sqrt{pq}}XM_{\bar{m}}X ,\label{inv2}
\end{eqnarray}
where $X$ is the Pauli-X matrix. The process of reversal is schematically represented in Fig. \ref{efig}. Either  $m$ or $\bar{m}$ is obtained after a measurement on $|\psi\rangle$. In the first case we apply $X$, measure, and if we obtain $m$ then  we can put the system back to the initial state $|\psi\rangle$ by applying another $X$ gate. In the case of the second occurrence, $\bar{m}$, we have a successful reversal only if we again get $\bar{m}$ when applying $X$ followed by a measurement; then we apply one more $X$ gate and the system goes back to the initial state. The probability of success is independent of the initial state in both situations. In the case of the upper path, the probability of obtaining the result $m$ is $P_{m}$; this has to be multiplied by the  (conditional) probability $P(m|m)$ of again getting the result $m$ after application of the gate $X$, which is given by
\begin{eqnarray}
P(m|m)&\stackrel{\rm def}{=}&\langle\psi_{m}|XM_{m}^{\dag}\times M_{m}X|\psi_{m}\rangle \nonumber \\
&=& P_{m}^{-1}\langle \psi |M_{m}^{\dag}XM_{m}^{\dag}X\times XM_{m}XM_{m} |\psi\rangle \nonumber\\
&=&(1-p)(1-q)P_{m}^{-1},\label{vreau1}
\end{eqnarray}
where for the last equality we used Eq. (\ref{inv1}).
Therefore, the total probability of success along the "$m$" path $|\psi\rangle\stackrel{m}{\rightarrow}|\psi\rangle$ is
$P_{|\psi\rangle\stackrel{m}{\rightarrow}|\psi\rangle} = P(m|m)P_{m} = (1-p)(1-q)$.
Similarly, for the "$\bar{m}$"path $|\psi\rangle\stackrel{\bar{m}}{\rightarrow}|\psi\rangle$ we get
\begin{eqnarray}
P(\bar{m}|\bar{m})&\stackrel{\rm def}{=}&\langle\psi_{m}|XM_{\bar{m}}^{\dag}\times M_{\bar{m}}X|\psi_{\bar{m}}\rangle \nonumber \\
&=& P_{\bar{m}}^{-1}\langle \psi |M_{\bar{m}}^{\dag}XM_{\bar {m}}^{\dag}X\times XM_{\bar{m}}XM_{\bar{m}} |\psi\rangle \nonumber\\
&=&pqP_{\bar{m}}^{-1}, \label{vreau2}
\end{eqnarray}
where the last equality follows from  Eq. (\ref{inv2}).
$P_{|\psi\rangle\stackrel{\bar{m}}{\rightarrow}|\psi\rangle} = P(\bar{m}|\bar{m})P_{\bar{m}} = pq$ is then  the probability of success for the path
$|\psi\rangle\stackrel{\bar{m}}{\rightarrow}|\psi\rangle$.


 We are now ready to address the problem of information extraction from a single qubit. Let start by considering precisely such a process (see Fig. \ref{efig}). At first sight it looks as if (with a certain probability of success), the unknown state of a single qubit can be determined by performing a series of measurements and reversing them. The probability of this happening is, admittedly, very small, but still finite. Suppose we have a total of $N$ successful reversals, out of which $N_{m}$ occurred {\it via} the upper-half paths $|\psi\rangle\stackrel{m}{\rightarrow}|\psi\rangle$ of the hexagons in Fig. \ref{efig}, and the other $N_{\bar{m}}=N-N_{m}$ occurred {\it via} the lower paths $|\psi\rangle\stackrel{\bar{m}}{\rightarrow}|\psi\rangle$. What we want is to estimate the state, that is, to find the two angles $\theta$ and $\varphi$ parametrizing any state $|\psi\rangle = \cos (\theta /2) |0\rangle +  \exp (i\varphi ) \sin (\theta /2) |1\rangle $ of a two-level system. To do so we use the maximum-likelihood estimator technique from statistical decision theory \cite{frieden}. We first notice that, in the Bayesian sense, for both paths there exist conditional probabilities ($P(m|m)$ and $P(\bar{m}|\bar{m})$) and priors ($P_{m}$ and $P_{\bar{m}}$). Thus we have to define the so-called weighted likelihood,
\begin{equation}
L (N_{m},N_{\bar{m}}) = [P_{m}]^{N_{m}} [P_{\bar{m}}]^{N_{\bar{m}}} [P (m|m)]^{N_{m}} [P(m|\bar{m})]^{N_{\bar{m}}}, \label{tr}
\end{equation}
which is the total probability (obtained as a product of probabilities of each event) that the chain of non-fail events in Fig. \ref{efig} has occurred. As before, to simplify the  notations we do not write explicitly the dependence on $(\theta ,\varphi)$, but we keep in mind that - as in standard decision theory -   that the likelihood $L (N_{m},N_{\bar{m}})$ is a probability density function on this two-parameter space.  Then we should find the "maximum {\it a posteriori} (MAP) estimate" \cite{frieden}, which in our case is the pair $(\theta ,\varphi)$ maximizing $L (N_{m},N_{\bar{m}})$, or, equivalently, $\ln L (N_{m},N_{\bar{m}})$. The next step would be to study how sensitive our measurement method is to variations of $(\theta ,\varphi )$ around their true value; this leads  to the concept of Fisher information. But this standard procedure does not lead anywhere, and the reason is that $L (N_{m},N_{\bar{m}})$ has in fact no dependence on $(\theta ,\varphi)$. Indeed, this can be seen immediately by noticing that the exponent of both $P_{m}$ and $P(m|m)$ (and $P_{\bar{m}}$ and $P(m|\bar{m})$ respectively) is the same (we know that the reversing procedure has been successful each time the result $m$ (respectively $\bar{m}$) has been obtained after a measurement), and by using Eqs. (\ref{vreau1}, \ref{vreau2}). This means that it is not possible to get any information about the state by the chain of measurements depicted in Fig. (\ref{efig}).

How can this be -  where has the information about switching into "$m$" or "$\bar{m}$" vanished? To understand what happens, let us take the logarithm of the weighted likelihood, which we call $-S = \ln L (N_{m},N_{\bar{m}})$; we then obtain
\begin{equation}
S = S_{{\rm meas}}  +  S_{{\rm rev}}
\end{equation}
where $S_{{\rm meas}} = -N_{m} \ln P_{m} - N_{\bar{m}} \ln P_{\bar{m}}$ represents the Shannon information obtained from the measurements, and
$S_{{\rm rev}} = -N_{m} \ln P(m|m) - N_{\bar{m}} \ln P(m|\bar{m})$ is the Shannon information resulting from the reversals. But again $P(m|m) = (1-p)(1-q) P_{m}^{-1}$ and
$P(\bar{m}|\bar{m}) = pqP_{\bar{m}}^{-1}$
 and, in the asymptotic approximation of a large number of events, $N_{m} = (1-p)(1-q) N$ and $N_{\bar{m}} = pq N$, therefore we have $S = -N [pq \ln pq + (1-p)(1-q) \ln (1-p)(1-q)]$. What happens is that the information resulting from reversal cancels exactly the information obtained via measurement (up to a constant). The remaining part is independent of the parameters $\theta$ and $\varphi$, and thus, overall, the measurement procedure from Fig. (\ref{efig}) is completely insensitive to the state parameters. One recognizes also that $S$ is the total conditional information (conditioned on the success of the reversal procedure) associated with the two paths, $S/N=-P_{|\psi\rangle\stackrel{m}{\rightarrow}|\psi\rangle} \ln
P_{|\psi\rangle\stackrel{m}{\rightarrow}|\psi\rangle}
-P_{|\psi\rangle\stackrel{\bar{m}}{\rightarrow}|\psi\rangle} \ln
P_{|\psi\rangle\stackrel{\bar{m}}{\rightarrow}|\psi\rangle}
$.
At $p=q=1/2$ this quantity reaches its maximum value of $\ln 2$. This entropy is all that remains after a chain of such events, representing the physical records of a string of $m$'s and $\bar{m}$'s (the information about which path the system actually took) that the experimentalist can write in the log notebook. The surprising fact is that, although all the measurements have been performed on a qubit prepared in a certain state, there is no information left in the environment about this state.  We also note that for the definition of standard estimation measures such as Fisher metric all possible results need to be accessible: the sum of the corresponding probabilities is 1, and the information thus defined is positive. But here we eliminate by postselection the situations in which the scheme fails, therefore we have to use conditional information, which generally is not guaranteed to be positive. In our case, the part containing  the $(\theta,\varphi)$-dependence is negative and exactly cancels $S_{{\rm meas}}$.  It simply has the meaning of an additional piece of information
 that logically contradicts some already-acquired knowledge. Finally, one can legitimately ask: what if we simply ignore the information coming from reversal? For example in Eq. (\ref{tr}) we write only the first two terms? The answer is that it can be done, but at one's own peril: then the value of $\theta$ obtained has no connection with the real one, and the procedure is in no way better than just guessing.

Finally, we point out that all the results derived above can be immediately generalized for any number of effects (although a simple physical implementation of such a measurement is not obvious). Define the measurement operators as
$M_{k}=\sqrt{q_{k}}|0\rangle\langle 0| + \sqrt{p_{k}}|0\rangle\langle 0|$ and  the corresponding effects $E_{k}=M_{k}^{\dag}M_{k}$, such that $\sum_{k} q_{k} = \sum_{k} p_{k} = 1$, therefore ensuring that $\sum_{k}E_{k} = I$. If none of  the $q_k$'s and $p_k$'s are zero, then for each result $k$ the measurement operator admits an inverse $M_{k}^{-1}=(p_{k}q_{k})^{-1/2}XM_{k}X$. Then in Fig. (\ref{efig}) we can have more then two paths (each indexed by $k$). The weighted likelihood Eq. (\ref{tr}) can be immediately generalized to this situation, and the proof of information cancellation along each path $k$ is similar.

We are now ready to address the following issue. Is the impossibility result above specific to the measurement scheme we have described, or do there exist more general physical principles from which it can be derived? In the following we discuss the relation between extraction of information from single quanta and the complementarity principle, the no-signaling principle, quantum teleportation, the no-cloning theorem, and quantum cryptography.

Take first the complementarity principle. Instead of using in  Eqs. (\ref{embarg},\ref{mg}) the basis $\{|0\rangle, |1\rangle\}$ (the eigenvectors of the Pauli-Z operator)  one can equally well use any other basis. For example the elements of the basis $|\pm\rangle = (|0\rangle \pm |1\rangle)/\sqrt{2}$ are eigenvectors of the Pauli-X operator. It is perfectly possible to have a series of measurements and reversals along $Z$, followed by a similar series along $X$. Still, it is not possible to claim that this is a joint measurement of two complementary observables, so no obvious contradiction is obtained.

Let us turn now to the EPR experiment. Suppose we have a maximally entangled Bell state between Alice's and Bob's qubits, $|\Phi^{+}\rangle = (1/\sqrt{2}) (|00\rangle + |11\rangle )$. Applying $M_{m}$ or $M_{\bar{m}}$ on Alice's qubit results in $(1/\sqrt{2-p-q})[\sqrt{1-q}|00\rangle + \sqrt{1-p}|11\rangle ]$ or in
$(1/\sqrt{p+q})[\sqrt{p}|00\rangle + \sqrt{q}|11\rangle ]$, which have concurrence \cite{concurrence}
$2\sqrt{(1-p)(1-q)}/(2-p-q)$ and $2\sqrt{pq}/{(p+q)}$ respectively. Both of these quantities are strictly smaller than 1 if $p\neq q$. Suppose now that we have obtained $m$ for the measurement on Alice's qubit. If $p$ is close enough to $1$ and $q\neq 1$, to a satisfactory good approximation we can claim that the state of Bob's qubit is $|0\rangle$. Now we reverse the measurement (because $p$ is large, the probability for succeeding is small but not zero). The interesting  thing that happens in this case is that we have restored the state $|\Phi^{+}\rangle$, {\it i.e.} we managed  to create a maximally entangled state from a state with almost zero entanglement. Thus partial measurements and their generalizations can be used to amplify entanglement!
 Furthermore, we can now perform a projective measurement in the $|\pm\rangle$ basis on Alice's qubit, which leaves Bob's qubit in the same state ($|+\rangle$ if Alice got + or $|-\rangle$ if Alice got -). It seems now that Alice can predict the values of the two noncommuting observables $Z$ and $X$ (the first to a controllable degree of approximation, the second exactly) of Bob's qubit! Unlike in  the original EPR argument where two sets of qubits are required for the argument, here this is achieved using only one pair.

Another important observation is that, if it were possible to determine the state of a single quantum object, the EPR pair could be used to signal faster then light: Alice could encode information as a direction in space and perform a von Neumann measurement along it. Bob is then left with a qubit oriented along the same direction,
and he can determine this state by using the measurement-and-reversal procedure. By {\it reductio ad absurdum}, extracting information from a single object is not possible.

Finally, let us consider the case of quantum teleportation. We show that, if it were possible to extract information from a single quantum object, then this scheme would allow
for remote cloning of a state using just two bits of information. This time, Alice has two qubits, the first in the unknown state  $|\psi \rangle = \cos (\theta /2) |0\rangle + \exp (i\varphi ) \sin (\theta /2) |1\rangle $,
and the second entangled with Bob's only qubit, in a Bell state $|\Phi^{+}\rangle$. Then Alice performs a controlled-not (CNOT) gate (on her second qubit conditioned on the first) followed by a Hadamard gate on the first qubit \cite{nielsen}. The result of this is
\begin{equation}
|\psi\rangle |\Phi^{+}\rangle \rightarrow \nonumber \\
\frac{1}{2} [|00\rangle|\psi\rangle
+  |01\rangle X|\psi\rangle +  |10\rangle Z|\psi\rangle +
 |11\rangle XZ|\psi\rangle ] \label{mini}
\end{equation}
Suppose now Alice is doing partial measurements on her qubits with strengths $q=0$ and $p$ close to 1, and the result of both measurements is $m$. Alice informs Bob about this, and Bob would be able to do a single-qubit partial-measurement tomography, which allows him to determine (within a certain degree of approximation and a lot of good luck) the state of his qubit. Note that a large amount of information can be encoded in the variables $\theta, \varphi$ (depending on the precision we want)
and that it took 2 bits of classical communication for Bob to be able to "decode" it.
 Moreover, even if only one bit of information can be communicated (corresponding to a measurement by Alice of her second qubit), Bob could still determine with arbitrary precision the value of $\theta$! From Eq. (\ref{mini}) it follows immediately that if Alice obtains 0, the resulting state is $\cos (\theta /2)|+\rangle|0\rangle + \exp (i\varphi ) \sin (\theta /2) |-\rangle|1\rangle$, while if she gets 1 the resulting state is $\cos(\theta /2) |+\rangle|1\rangle + \exp (i \varphi ) \sin (\theta /2) |-\rangle|0\rangle$. Bob now measures his probabilities of obtaining $0$ and $1$ and he uses the classically-transmitted bit of  information to decide which one of these probabilities to associate with the amplitudes $\cos(\theta /2)$ and $\exp (i \varphi ) \sin (\theta /2)$. Moreover, Alice can also in principle recover her state exactly: as a result, Bob ends up again with a qubit maximally entangled with Alice's second qubit, but also with some classical information about the qubit which in principle could allow him to built another qubit in approximately the same state. Note that this procedure would not contradict directly the no-cloning theorem (which is proved using only unitary transformations), and it would allow us to built a relatively simple probabilistic cloning machine. However, it is still forbidden by quantum mechanics, as we have shown above -- to get any true information, one needs to have an ensemble. Somehow, quantum physics does not like to provide all the information at once. Much like a hero in a treasure-hunting novel, Bob gets one little clue at a time (each time he measures an element of the ensemble).

Let us now examine the problem of quantum key distribution. We want to show that if extracting information from a single qubit were possible, this would provide a hacking strategy for quantum key distribution protocols. Take for example the B92 protocol \cite{nielsen}. Alice generates a random string $\{a\}=\{0,1\}$ of classical bits, and if $a=0$ she sends the qubit $|0\rangle$ to Bob, while if $a=1$ she send $|+\rangle$. Bob generates his own random classical bits $\{a'\}$  and measures $Z$ on the qubit sent by Alice if $a'=0$ (in which case he obtains the result -1 only if $a=1$), or $X$ if $a'=1$ (in which case he obtains the result -1 only if $a=0$). After discussing over a classical channel, Alice and Bob
keep only the qubits for which the result -1 has been obtained  -- the corresponding classical bits $\{a\}$ and $\{a'\}$ will be anticorrelated $a = 1-a'$ and a shared secret key is obtained. If, however, Eve intercepts the qubit she could perform an approximate partial-measurement quantum tomography. If she fails, she does nothing and Bob will interpret the result as the qubit being lost in the communication channel; if she succeeds, she learns approximately the state of the qubit (the value of $a$), and she can forward the qubit to Bob.
Together with  the value of Bob's result (which is publicly broadcasted), she could infer the value of $a'$, that is she would find the key shared by Alice and Bob without any of them noticing.

Finally, if there is no element of reality for {\it wavefunctions}, perhaps there could be one for {\it entanglement} \cite{zeilinger}? Suppose we are interested in two-qubit systems and we use an ancilla as a probe. If instead of an ensemble we have just two qubits, can we measure their entanglement? The results above show that the answer is negative. One still has to erase all the classical information in order to reverse the measurements.

  In conclusion, we proved that it is not possible to extract even probabilistically any information from a single qubit prepared in an unknown state by using a generalized version of partial measurements. We also examine how this result is connected to general principles such as no-signaling, no-cloning, complementarity, the possibility of quantum teleportation, and the security of quantum key distribution.

\acknowledgements
Financial support from the Academy of Finland is acknowledged (grant 00857, and projects 129896, 118122, and 135135).

\end{document}